\documentclass[english]{article}
\usepackage[T1]{fontenc}
\usepackage[latin9]{inputenc}
\usepackage{babel}
\usepackage{amsmath,amsfonts}
\usepackage{amsthm}
\usepackage{graphicx}
\usepackage{comment}
\usepackage[authoryear]{natbib}
\usepackage[unicode=true,pdfusetitle,
 bookmarks=true,bookmarksnumbered=false,bookmarksopen=false,
 breaklinks=false,pdfborder={0 0 0},pdfborderstyle={},backref=false,colorlinks=false]
 {hyperref}
\usepackage{authblk} 

\makeatletter
\providecommand{\theoremname}{Theorem}
\theoremstyle{plain}
\newtheorem{thm}{\protect\theoremname}

\usepackage{amsmath}
\usepackage{fullpage}

\usepackage{bbm}

\@ifundefined{showcaptionsetup}{}{%
 \PassOptionsToPackage{caption=false}{subfig}}
\usepackage{subfig}
\makeatother

\begin{document}
\title{With big data come big problems:\\
pitfalls in measuring basis risk for crop index insurance}
\author[1]{Matthieu Stigler}
\author[2]{Apratim Dey}
\author[3]{Andrew Hobbs}
\author[1]{David Lobell}

\affil[1]{Center on Food Security and the Environment, Stanford University, USA}
\affil[2]{Department of Statistics, Stanford University, USA}
\affil[3]{Department of Economics, University of San Francisco, USA}

\date{November 30, 2021}

\maketitle
New satellite sensors will soon make it possible to estimate field-level crop yields, showing a great potential for agricultural index insurance. This paper identifies an important threat to better insurance from these new technologies: data with many fields and few years can yield downward biased estimates of basis risk, a fundamental metric in index insurance. To demonstrate this bias, we use state-of-the-art satellite-based data on agricultural yields in the US and in Kenya to estimate and simulate basis risk. We find a substantive downward bias leading to a systematic overestimation of insurance quality.
 
In this paper, we argue that big data in crop insurance can lead to a new situation where the number of variables $N$ largely exceeds the number of observations $T$. In such a situation where $T\ll N$, conventional asymptotics break, as evidenced by the large bias we find in simulations. We show how the high-dimension, low-sample-size (HDLSS) asymptotics, together with the spiked covariance model, provide a more relevant framework for the $T\ll N$ case encountered in index insurance. More precisely, we derive the asymptotic distribution of the relative share of the first eigenvalue of the covariance matrix, a measure of systematic risk in index insurance. Our formula accurately approximates the empirical bias simulated from the satellite data, and provides a useful tool for practitioners to quantify bias in insurance quality.

\vspace{1cm}

\section{Introduction}

Index insurance is a promising tool to reduce the risk faced by smallholder farmers. By linking payouts to a regional index instead of individual losses, it reduces moral hazard, adverse selection and transaction costs. But delinking payouts from individual losses creates \emph{basis risk}, the possibility that a farmer experiences a loss yet does not receive any indemnity. Basis risk substantially reduces the benefit of index insurance, and if severe enough can make it worse than no insurance at all. Further, \cite{Clarke2016}, \cite{carter2017index}, and others argue that basis risk is likely among the most important barriers to index insurance adoption, and that the basis risk of some index insurance schemes has been be very high \citep{clarke2012weather,JensenBarrettEtAl2016}. 

Recent improvements in satellite remote sensing and machine learning show great potential to improve the accuracy of index and thus reduce basis risk. These technologies have triggered a very active literature extending far beyond the field of economics.\footnote{See for example the special issue of \textit{Remote Sensing} on ``Earth Observation for Index Insurance''  2021, 13(5), or reviews by \citet{BenamiJinEtAl2021} and \citet{DeLeeuwVrielingEtAl2014}.}  These technologies have led to three major shifts in the design of agricultural indices. First, satellite data has helped to design better weather-based indices: while early products were based on local weather stations, satellite data has increased the spatial resolution of weather indices, and facilitated the incorporation of new variables such as soil moisture. Second, it has led to a shift from input-based weather indices towards higher-accuracy output-based indices based on vegetation indices observed with optical satellite sensors. In these two first approaches, the satellite data is used primarily to obtain a better index, while the assessment of the quality of the index is conducted using traditionally-collected field-level yield data. In the third approach, satellite data is used directly to estimate farm or field-level crop yields. Although accurately predicting individual yields currently remains a challenge, rapid progress is being made and this approach shows great potential for deriving very accurate output-based indices. Even more importantly, it will provide a cost-effective way to assess the quality of a given index over a large number of fields, helping insurers to design better insurance zones and making it easier for governments, researchers, and others to reliably assess the quality of index insurance products.

This paper identifies and studies a new challenge associated with assessing index insurance quality with more granular data: existing measures of quality are biased in large-$N$ (number of farms), small-$T$ (number of time periods) samples. We show that $R^2$-derived estimates of basis risk are systematically biased downward in the small-$T$ large-$N$ case, meaning practitioners who do not take this into account will generally overestimate the quality of index insurance products. Intuitively, this bias arises from the fact that the basis risk estimates are functions of the covariance matrix between fields. The covariance matrix between fields has $N\cdot(N-1)/2$ parameters, yet only $N\cdot T$ observations to estimate it. Having more fields $N$ than time periods $T$ is very typical of agricultural data, and is going to be exacerbated by developments of satellite data methods, which are particularly suited to extend the sample over space, 
but are unfortunately only available for a few recent years. The resulting bias has gone unnoticed in the existing literature and has important implications for both past and future estimates of basis risk.

After documenting the bias in various measures of basis risk, we analyze it theoretically. We focus on linear measures of basis risk, which allows us to connect our problem with a rich literature in statistics. We start with a review of models to parameterize the inter-field covariance matrix, focusing on the \emph{spiked} model introduced by \citet{Johnstone2001}. We discuss then how the high-dimension, low-sample-size (HDLSS) framework introduced by \citet{HallMarronEtAl2005} can help us understand the bias of linear basis risk measures. In the HDLSS framework, the sample size $T$ is assumed fixed, while the number $N$ of variables is assumed to grow to infinity. This corresponds exactly to the situation we are facing in index insurance, where satellite data techniques increase sample sizes over space much faster than over time. Using results from the HDLSS literature, we provide a new theorem deriving the theoretical bias of our linear measure of basis risk. Going back to the simulations, we find that our theory predicts the empirical bias remarkably well. 

The bias we study in this paper is particularly pernicious for two reasons. First, we show the bias in basis risk measurement can actually be worsened by higher resolution data. Second, the bias is greatest when individual yields are poorly correlated, meaning it is likely to be particularly severe in smallholder systems in developing countries, which is precisely where new satellite data promise to have the largest positive impact. In light of these findings, understanding this bias is essential to realizing the promise of high resolution satellite data for index insurance.

This paper's findings also provide a potential explanation for observed low uptake of existing index insurance products.\footnote{See \cite{carter2014index} and \cite{carter2017index} for discussions of the literature on low index insurance uptake} Farmers are experts on their yields and are aware of how they relate to their neighbors, and likely have an accurate understanding of how correlated their yields are to their neighbors'. Since the bias we study in this paper is essentially a result of inaccurate estimates of inter-field correlations due to a small number of observed time periods, farmers who have more accurate understandings of these correlations ought to buy insurance less often than biased estimates would suggest.

\section{Measures of basis risk}

Index insurance products are usually assessed following two broad approaches. In the first one, the interest is on basis risk, which is a measure of the frequency and size of errors in predicting individual yields and/or harvest losses. The second seeks to evaluate how the insurance product derived from this index performs. This entails typically specifying a indemnity and premium functions and assuming a utility function for the farmer. In this paper, we focus on the first approach. Basis risk is sometimes defined as the probability of a farmer experiencing a loss yet not receiving an indemnity. As such, it can be estimated as a simple conditional probability, the false negative probability. However, this measure fails to capture the severity of the prediction error, which has important consequences for farmer welfare. For this reason, \citet{ElabedBellemareEtAl2013} suggest focusing on the field-level share of the variation in yields not explained by the index. This is equivalent to using $1-R^2_i$, where $R^2_i$ is the coefficient of determination between the yields of field $i$ and the index.

In this paper, we focus on output-based indices such as the zone average yield. This comes from our initial motivation to assess the benefits of third-generation datasets, which (will) allow estimates of yields at the field-level. That said, the same results hold for traditional area yield insurance contracts; the source of the field-level yield estimates does not matter.  In recent work on output-based indices, \citet{StiglerLobellKenyaOptimal2021} discuss how to aggregate the field-specific $R_{i}^{2}$ and propose a zone-specific \emph{total} $\overline{\overline{R^{2}}}\equiv1-\sum_{i}SSR_{i}/\sum SST_{i}$. This $\overline{\overline{R^{2}}}$ is a generalization of the individual $R^{2}$ written as $R_{i}^{2}\equiv1-SSR_{i}/SST_{i}$, where SSR and SST stand respectively for sum of squared residuals and total sum of squares. This measure is simply a variance-weighted average of the individual $R_i^2$, meaning it puts more weight on farmers who are exposed to more risk. The $\overline{\overline{R^{2}}}$  measure can alternatively  be obtained by running $N$ field-specific regressions of the index on individual yields and aggregating their $SSR_i$ and $SST_i$ into  $\overline{\overline{R^{2}}}\equiv1-\sum_{i}SSR_{i}/\sum SST_{i}$. In the case of an output-based area-yield index, \citet{StiglerLobellKenyaOptimal2021} propose an alternate, numerically identical, formula:

\begin{equation}\label{eq:R2_mean}
\overline{\overline{R^{2}}}=tr\left(\Sigma\mathbf{1}(\mathbf{1}^{'}\Sigma\mathbf{1})^{-1}\mathbf{1}^{'}\Sigma\right)/tr(\Sigma)
\end{equation}

Here, $\Sigma$ is the covariance matrix between individual fields, and $\mathbf{1}$ denotes a vector of 1, of dimension $N$. While numerically equivalent to field-specific regression, formula \eqref{eq:R2_mean} has the advantage of establishing the connection between the basis risk and the covariance of fields.  Intuitively, the strength of the index depends on the strength of the off-diagonal elements: a diagonal covariance matrix (uncorrelated fields) would result in higher basis risk than a covariance matrix with many positive off-diagonal elements (many correlated fields).  

\citet{StiglerLobellKenyaOptimal2021} show also that the formula \eqref{eq:R2_mean} can be generalized to a broader class of output-based indices, which use field-specific weights to form the index, $f_{t}=\sum_{i}w_{i}y_{it}$, or in matrix form, $f=Y\mathbf{w}$. The area-yield index is a special case in this class, with $\mathbf{w}=\mathbf{1}/N$. The formula becomes then:

\begin{equation}\label{eq:R2_w}
\overline{\overline{R^{2}(\mathbf{w})}}=tr\left(\Sigma\mathbf{w}(\mathbf{w}^{'}\Sigma\mathbf{w})^{-1}\mathbf{w}^{'}\Sigma\right)/tr(\Sigma)
\end{equation}

They show that this quantity is not maximized using the area-yield index $\mathbf{1}/N$, but instead taking the first principal component (PC) of the covariance matrix $\Sigma$, $\mathbf{w^\star}=PC_1(\Sigma)$. Evaluated at this optimal $\mathbf{w}^{\star}$,
the objective function $\overline{\overline{R^{2}(\mathbf{w}^{\star})}}$ turns out to be equal to the share of the first eigenvalue of $\Sigma$, that is $\overline{\overline{R^{2}(\mathbf{w}^{\star})}}=\lambda_{1}/\sum\lambda$. The $\overline{\overline{R^{2}(\mathbf{w}^{\star})}}$ is an interesting measure that defines the upper-bound any index can achieve (according to the total $R^{2}$ criterion) for a given zone. In that sense, it can be interpreted as a measure of zone quality, a low $\overline{\overline{R^{2}(\mathbf{w}^{\star})}}$ for a given zone indicating that even the best index would not perform very well. The connection between $\overline{\overline{R^{2}(\mathbf{w}^{\star})}}$ and the eigenvalues of $\Sigma$ indicates that the $\overline{\overline{R^{2}(\mathbf{w}^{\star})}}$ is equivalent to the usual definition of $\lambda_{1}/\sum\lambda$ in terms of the \emph{percentage of total variance captured by the first principal component}. In addition, and of particular relevance for the current paper, the statistical properties of sample eigenvalues are a very well-studied problem in statistics. 

Admittedly, linear correlation measures have limitations in the context of index insurance. Arguably, it is more important for an index to accurately predict yield losses than to predict good harvests. Various approaches have been suggested to take this into account, ranging from quantile regression \citep{ConradtFingerEtAl2015}  to more sophisticated left-tail dependence indices \citep{Bokusheva2018}. In the following, we include also a quantile version of our total $R^2$ measure, based on the quantile pseudo $R^2$ developed by \citet{KoenkerMachado1999}. \citet{KoenkerMachado1999} suggest a pseudo $R^2(\tau)=1-V(f,\tau)/V(const,\tau)$ at quantile $\tau$, where $V(\tau)$ is the quantile analogous to the SSR. In a similar way to our total $R^2$, we define the total quantile pseudo $R^2$ as $R^2_q=1 -\sum V_i(f,\tau)/\sum V_i(const,\tau)$, and use the value of $\tau=0.3$ following previous literature \citep{BucheliDalhausEtAl2020}. The bias we identify in this study may also apply to other nonlinear measures of basis risk, but we leave that for future studies.

\section{Data and empirical simulations\label{sec:Data-and-empirical-simuls}}

\subsection{Data}

We use two state-of-the-art datasets of satellite-estimated yields in the USA and in Kenya to illustrate the potential bias in basis risk measures. Both datasets contain maize yield predictions produced with the Scalable Yield Mapper (SCYM) model initially developed by \citet{LobellEtAl2015}. The SCYM model is one of the most advanced yield prediction models available to date (see \citealp{JinAzzariEtAl2017, AzzariLobell2017, DeinesPatelEtAl2021OneMillionTruth} for the US and \citealp{BurkeLobell2016, JinAzzariEtAl2017SmallholderYieldHeterogeneity, JinAzzariEtAl2019, LobellAzzariEtAl2020AJAE}  for Sub-Saharan Africa), and has been already used to analyze various questions such as the effect of cover crops, of conservation tillage or the dynamics of crop expansion \citep{SeifertAzzariEtAl2018, DeinesWangEtAl2019, Stigler2018}. The dataset has been used specifically for 
analyzing crop insurance in the US in \citet{StiglerLobellUSIns2020} and in Kenya in \citet{StiglerLobellKenyaOptimal2021}.

While the US and Kenya datasets share a common methodology, they also differ in several respects. First, we have twenty years of data for the US: from 2000 to 2019, while for Kenya we have only four: from 2016 to 2019. This difference is due to the fact that fields are much smaller in Kenya which means that higher resolution satellite images are required (from Sentinel-2), and those images are only available for recent years. Another  difference lies in the fact that maize is mainly cultivated in rotation together with soybeans in the US, while this is less common in Kenya. For the US, this means we observe a large number of missing maize values for those fields practicing rotation, which makes estimating accurate covariance matrices more difficult. We adopt a simple solution, and focus on the fields that only cultivate corn over the 2000-2019 period, and select counties that have at least 30 observations.  Doing so, we end up with a sample of 37 \emph{zones} for the US. For Kenya, almost all fields cultivate maize every year, and we randomly sample 200 fields in each of 453 Kenyan sub-counties (Kenya's smallest administrative unit). A last aspect where both datasets differ is in the quality of the satellite predictions. Predictions are typically much better in the US, characterized by large uniform fields, than in Kenya, which has smaller fields and more heterogeneous cultivation practices. \citet{DeinesPatelEtAl2021OneMillionTruth} report that the SCYM yield estimates in the US have a $r^2$ of 0.45 when compared to a ground-truth dataset at the field level, raising to 0.69 when assessed against county-level means instead. On the other hand, \citet{JinAzzariEtAl2019} report that the yield estimates in Kenya have an agreement of about 50\% against district means. Clearly, the current accuracy of the satellite-based yield predictions is not yet perfect, and more research is still needed before using these datasets at an operational level for routine insurance assessment. Witnessing the rapid progress made in the field in the last ten years, there is good reason to believe that in a near future yield estimates will be much more accurate. In this paper, we focus on another source of error for insurance applications, that due to the small-$T$ large-$N$ setting, that has largely gone unnoticed in the literature. 

\subsection{Basis risk measures in SCYM data}

Our analysis here proceeds in two steps. We first estimate various basis risk measures using both SCYM data sets. We then estimate the bias associated to these measures. To do so, we employ a Monte Carlo (MC) simulation experiment using pseudo true values calibrated to the data at hand. More precisely, we pretend that the covariance matrices and the basis risk measures estimated in the first step are the true ones, and simulate random samples of various $T$ sizes using these pseudo-true covariances. We then re-estimate the basis risk measures on the simulated samples, and infer the bias by comparing those simulated values to the pseudo-true basis risk measures used to simulate the data. 

\begin{figure}
  \caption{Basis risk measure on SCYM data\label{fig:Basis-risk-SCYM-empirial}}
  \includegraphics[width=0.95\columnwidth]{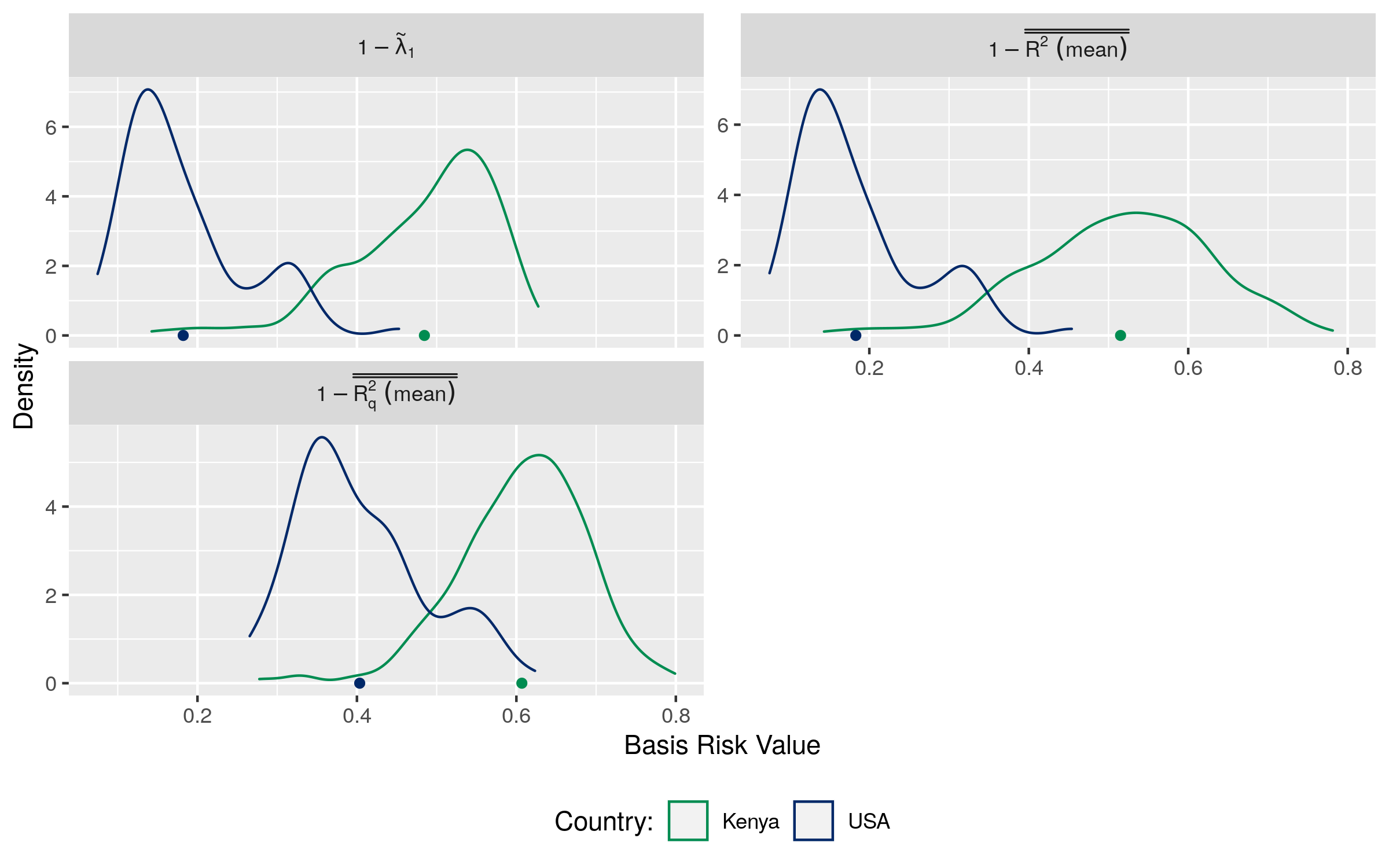}
\end{figure}

Starting with the initial estimates of basis risk, we focus on three measures, 1) the total $R^2$ using the county mean as the index, 2) the total quantile pseudo $R^2$ using also the county mean, and 3) the total $R^2$ using the optimal index, which is equivalent to the share of the first eigenvalue. Figure~\ref{fig:Basis-risk-SCYM-empirial} shows the results for these three measures, highlighting a stark difference between Kenya and the USA. Remembering that basis risk is defined as $1-R^2$, we see that the basis risk is much lower  in the USA than in Kenya according to the three measures.\footnote{Interestingly, these three measures of basis risk are highly correlated to each other, the lowest correlation being 92\% between the quantile pseudo $R^2$ and the first eigenvalue.} This result is consistent with the structure of agricultural production in the two settings, large-scale farms in the US use relatively similar production technologies, whereas smallholder farmers Kenya are very heterogeneous. It is therefore unsurprising that basis risk, interpreted as the lack of correlation between fields, is much higher in Kenya than in the USA. However, these estimates are based on small-T samples (T=4 for Kenya, and T=20 for the USA), and we show in the next section that they underestimate basis risk as predicted by the theory.

\subsection{SCYM-based simulations of the bias}

To gauge the reliability of the estimates obtained above, we proceed to a Monte Carlo simulation. Using the covariance matrices and means implicitly estimated above and pretending those are pseudo-true covariance matrices, we simulate random samples assuming a normal distribution. We do this for three different T dimensions representative of dataset found in practice, T=4, 10 and 20, and for each of these run 500 simulations. For each simulated sample, we recompute the three measures of basis risk, average them over the 500 samples, and compute the resulting bias by comparing these averages to the pseudo-true values obtained from the pseudo-true covariance matrices.\footnote{For the linear basis risk metrics, the population value is directly obtained from on the covariance matrix based on the formulas derived above. For other measures such as the quantile pseudo $R^2$, we don't have analytical formulas for the population values, and hence obtain them by simulating with a large sample of 25000 observations.} 

\begin{figure}
\caption{Bias from simulations based on SCYM\label{fig:bias-SCYM-simul}}

\includegraphics[width=0.95\columnwidth]{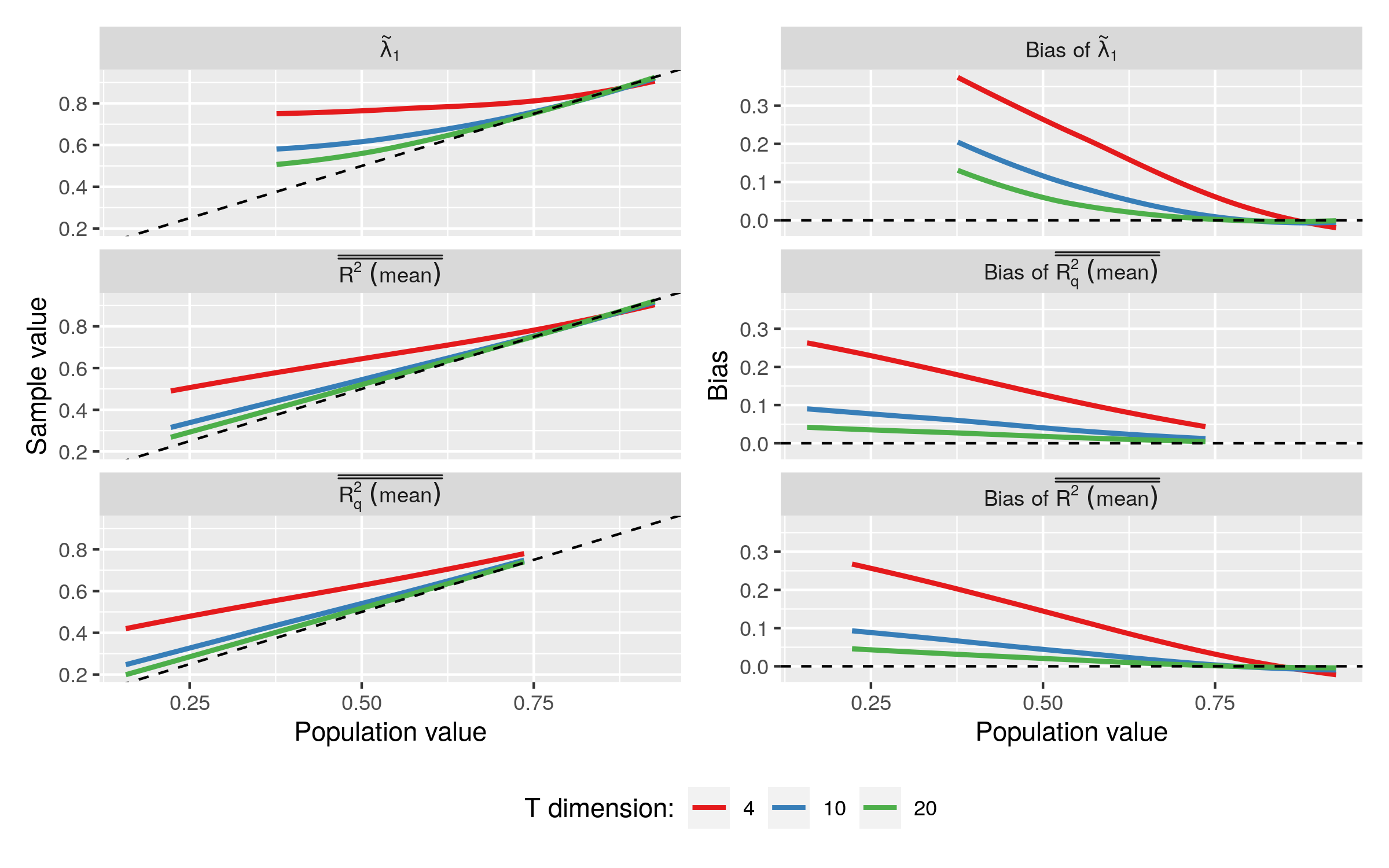}
\end{figure}

Figure~\ref{fig:bias-SCYM-simul} shows the bias as estimated by simulation. Values on the x-axis denote the pseudo-true population value, and on the y-axis the average sample values (left column) and bias (right column). We note a very similar phenomenon over each basis risk metric: the bias is relatively high for low values of the pseudo-true basis risk, and tend to decrease for increasing values of basis risk, with a possible sign reversal for very high values. This indicates that an insurer assessing the basis risk of a zone will tend to be over-optimistic about the $R^2$ measure, and hence under-estimate the basis risk itself. This upward bias in the $R^2$ is relatively large, and even larger in percentage terms, with an upward bias of 150\% for low values of the quantile pseudo $R^2$ measure. The bias decreases for larger values of T, suggesting it would eventually disappear with a large enough sample over the T dimension. Taken together, the fact that the bias is higher for lower value and for lower T suggests that the bias in the initial estimates in Figure~\ref{fig:Basis-risk-SCYM-empirial} is relatively modest for the USA (T=20 and high $R^2$ values observed), but much more important in Kenya (T=4 and low $R^2$ values observed).

The calibrated Monte Carlo exercise suggests that there is a substantive upwards bias in the total $R^2$ measure, resulting in downwards bias for the linear basis risk metrics considered. However, this approach of simulating pseudo-true covariance matrices faces several limitations. First, it rests on an initial estimate of the covariance matrix, which is itself potentially biased. Second, and most important, it is difficult to know whether these insights generalize to other cases, since the results were derived for specific covariance matrices. Generalizing these results calls for analyzing the bias theoretically, which is the subject of the next section.





\section{Theory\label{sec:Theoretical-simulations}}

The simulations in Section~\ref{sec:Data-and-empirical-simuls} show that there is potentially a substantive bias in various basis risk measures. These results were obtained by pretending that the sample covariance matrices we estimated are the population ones, and simulating from this population matrix. This is somewhat artificial, since the simulations show that our initial estimate could be biased. Escaping this conundrum calls for a more formal approach, assuming a-priori a specific covariance matrix to be used as data-generating process (DGP). This makes it possible to simulate data by controlling and varying the \emph{true} parameters, and assessing how sample estimates behave with respect to these known parameters.

\subsection{Choice of the data generating process}

To derive analytical results on the estimation of the basis risk metrics, we need to specify a data generating process. To do so, we adopt the so-called spiked covariance model introduced by \citet{Johnstone2001}. This model assumes that the eigenvalues of the covariance matrix are spiked, i.e. that a few eigenvalues clearly dominate the rest of the eigenvalues. This corresponds to the idea that a few common factors explain much of the variability of the data, and that the remaining variability is idiosyncratic. In fact, \citet{WangFan2017} show that a spiked structure can be generated from a factor model. This factor structure is exactly the assumption behind index insurance, which posits that a single common index will predict well individual yields. It is also implicitly assumed in existing literature on index insurance that models farm-level risk as a linear combination of covariate shocks and idiosyncratic farm-level shocks (See \citealt{Miranda1991}, \citealt{Mahul1999} and \citealt{ConradtFingerEtAl2015} for example). This suggests that the spiked model is a very natural starting point as generating process for agricultural data, noting that the main determinants of yield (weather, input and output prices) are highly correlated within a zone. As another benefit, the spiked model is the subject of a vast theoretical literature (see for reviews \citealp{JohnstonePaul2018}), providing us with important tools to understand the bias of our estimates. 

The idea behind the spiked model is to specify the eigenvalues of the covariance matrix, assuming that a few eigenvalues are much larger than the other ones, and furthermore grow with the dimensionality $N$ of the sample. In other words, letting $\lambda_{1}>\lambda_{2}>\ldots>\lambda_{M}>\lambda_{M_{1}}\geq\ldots\geq\lambda_{N}$
represent the ordered eigenvalues of $\Sigma$, one assumes that there are at most $M$ \emph{large} eigenvalues, and that the remaining $N-M$ are bounded. In the following, we will use the simplest specification, assuming that there is only one large spike $\lambda_{1}=aN^{\alpha}$, and that the remaining eigenvalues are all equal to a constant $b$. Denoting $\Lambda$ the diagonal matrix whose elements are the ordered eigenvalues, the population covariance matrix is then specified as $\Sigma=Q\Lambda Q^{'}$, where $Q$ is an orthogonal matrix ($QQ^{'}=I$). Using an orthogonal matrix guarantees that the eigenvalues of $\Sigma$ are the same as $\Lambda$, that is $[aN^{\alpha},b,\ldots,b]$. 

As discussed above, a measure of minimum zonal risk is given by $1-\overline{\overline{R^{2}(w^{\star})}}=1-\lambda_{1}/\sum\lambda\equiv1-\tilde{\lambda}_{1}$,
which indicates the minimum $1-\overline{\overline{R^{2}(w)}}$ that can be reached within a zone by any linear index. From now on, we will focus on the population $\tilde{\lambda}_{1}\equiv\lambda_{1}/\sum\lambda$, and its sample counterpart, $\hat{\tilde{\lambda}}_{1}\equiv\hat{\lambda}_{1}/\sum\hat{\lambda}$. Starting with the population $\tilde{\lambda}_{1}$, in the spiked model it takes the value of $\tilde{\lambda}_{1}=aN^{\alpha}/(aN^{\alpha}+(N-1)b)$. The behavior of $\tilde{\lambda}_{1}$ with growing $N$ will depend on the value of $\alpha$, and we distinguish three cases:

\begin{itemize}
\item Vanishing spike: $\alpha<1$, and hence $\tilde{\lambda}_{1}\xrightarrow{N\to\infty}0$ 
\item Constant spike: $\alpha=1$, and hence $\tilde{\lambda}_{1}\xrightarrow{N\to\infty}a/(a+b)$ 
\item Expanding spike: $\alpha>1$, and hence $\tilde{\lambda}_{1}\xrightarrow{N\to\infty}1$ 
\end{itemize}

Which spike structure should be considered is a complicated question. Typically, a researcher faces a dataset with given $T$ and $N$, and asking the thought experiment of what would happen if she had an infinite number of fields N is somewhat abstract. One could eventually argue that when the dimension $N$ is increased by extending the sample over space, adding fields further away would possibly reduce the value of $\tilde{\lambda}_{1}$, which would correspond to the vanishing spike case. Conversely, if the sample is extended by adding more pixels or fields within a zone, $\tilde{\lambda}_{1}$ could alternatively increase (expanding spike). At the theoretical level, the vanishing and expanding spike are, however, not very interesting since they only allow two extreme values of either 0 or 1. In the following, we focus hence on the constant spike, assuming that the value of $\tilde{\lambda}_{1}$ is constant, and equal to $a/(a+b)$. This will allow us to investigate the behavior of the basis risk measure for a variety of cases of $\tilde{\lambda}_{1}$, representing both low and high homogeneity zones that we encountered in Section~\ref{sec:Data-and-empirical-simuls}. 

\subsection{Analytical results}

Turning now to the behavior of the sample eigenvalue under the moderate spike, we need now to make assumptions on the ratio between $T$ and $N$. There exist broadly three frameworks in statistics: 1) traditional asymptotics, with $N$ fixed and $T\to\infty$ so that $N/T\to0$, 2) random matrix theory $N/T\to c$ for a constant $c
$ and 3) high-dimension low-sample size (HDLSS), where $N/T\to\infty$. The latter case, HDLSS, appears the most appropriate to describe third-generation datasets, where $T$ is considered fixed, and $N$ is allowed to grow very large. The HDLSS was introduced by \citet{HallMarronEtAl2005}, with notable contributions from \citet{AhnMarronEtAl2007,JungMarron2009}, see also \citet{AoshimaShenEtAl2018} for a review. While important results have been derived for the raw sample eigenvalue $\hat{\lambda}_{1}$ (see \citealp{AhnMarronEtAl2007}), there is, to the best of our knowledge, no result available on the relative eigenvalue $\hat{\tilde{\lambda}}_{1}\equiv\hat{\lambda}_{1}/\sum\hat{\lambda}$. To fill this gap, we derive a new result, describing the distribution and the bias of $\hat{\tilde{\lambda}}_{1}$, building on the seminal work by \citet{AhnMarronEtAl2007}. In what follows, $\stackrel{d}{\to}$ denotes convergence in distribution and $\stackrel{p}{\to}$ denotes convergence in probability. See the appendix for details on the notation.

\begin{thm}[Distribution of the share of the first sample eigenvalue.]
\label{thm:main-theorem-sample-eigenval}

For $1\leq t\leq T$, set $Y_t=(y_{1t},...,y_{Nt})$ and assume that $ Y_t\stackrel{iid}{\sim}\mathcal N_N( \mu,\Sigma)$ where $\mathcal N_N( \mu,\Sigma)$ denotes the $N$-dimensional Normal distribution with mean vector $ \mu\in\mathbb R^N$ and covariance matrix $\Sigma\in\mathbb R^{N\times N}$, and $mu$ is unknown. Also assume a spiked model for $\Sigma$ i.e. the eigenvalues of $\Sigma$ satisfy $\lambda_1=aN^\alpha,$ and $\lambda_2=\cdots=\lambda_N=b$ for some positive constants $a,b$. We have the following results.
\begin{enumerate}
\item Vanishing spike:  $\hat{\tilde{\lambda}}_1 \stackrel{p}{\to} \tfrac{1}{T-1} $, noting that $\tilde{\lambda}_1 \to 0 $.
\item Constant spike: $\hat{\tilde{\lambda}}_1 \stackrel{d}{\to}\frac{aC^2+b}{aC^2+b(T-1)} $ where $C^2\sim\chi^2_{T-1}$, noting that $\tilde{\lambda}_1\to\dfrac{a}{a+b}$.
\item Expanding spike:  $\hat{\tilde{\lambda}}_1 \stackrel{p}{\to} 1 $ noting that $\tilde{\lambda}_1 \to 1 $.
\end{enumerate}
\end{thm}

\begin{thm}[Bias of the share of the first sample eigenvalue.]
\label{thm:theorem-bias}
Assume the same setup as in Theorem \ref{thm:main-theorem-sample-eigenval}.
\begin{enumerate}
\item Vanishing spike: Asymptotic bias is $\tfrac{1}{T-1}$.
\item Constant spike: Let $r=\tfrac{a}{a+b}$. Asymptotic bias is $\mathbb E\left(\tfrac{(1-r)(rC^2+1-r(T-1))}{rC^2+(1-r)(T-1)}\right)$ where $C^2\sim\chi^2_{T-1}$.
\item Expanding spike: Asymptotic bias is 0. 
\end{enumerate}
\end{thm}

\begin{figure}
  \caption{Asymptotic distribution of $\hat{\tilde{\lambda}}_{1}$ \label{fig:distribution-lambda}}

  \begin{centering}
    \includegraphics[width=0.95\columnwidth]{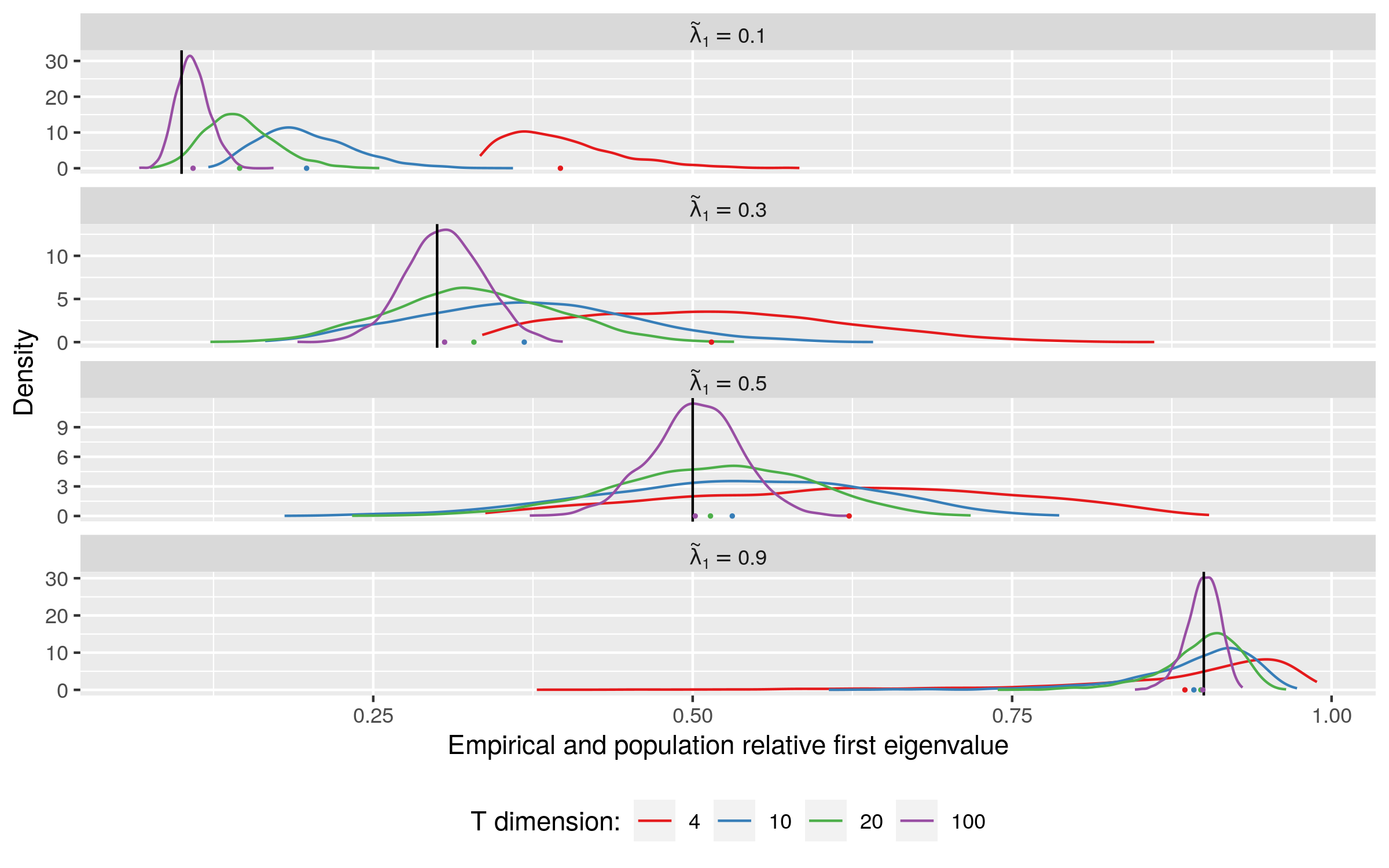}
  \end{centering}
\end{figure}

Figure~\ref{fig:distribution-lambda} shows the asymptotic distribution of $\hat{\tilde{\lambda}}_{1}$ in the constant spike case for four values of the true $\tilde{\lambda}_{1}$ and for various T dimensions. The asymptotic distribution explains the behavior of the bias that we observed in the calibrated simulations above. It is clear that the bias is large for low values of the true $\tilde{\lambda}_{1}$, and tends to decrease as the true $\tilde{\lambda}_{1}$ increases. This is especially problematic: the bias is greater when there is little shared risk between farmers. This is likely to be the case in places where farmers are more heterogeneous, such as smallholder systems in developing countries.  It is also clear that the bias is a function of $T$: increasing $T$ reduces the bias, and for values as high as $T=100$, this bias appears negligible. Unfortunately, having 100 years of field-level data ($T=100$) is unrealistic, and climate change would limit the usefulness of such a long series even if it were available. Looking at a value of $T=20$ which is more realistic for agricultural data, bias is still present, in particular for lower values of $\tilde{\lambda}_{1}$. This suggests that even with high-quality data, an insurer will still be over-confident in her assessment of the quality of an insurance product, unless the true quality is very high. 

Ideally, deriving the bias would help construct a bias-corrected estimator. Unfortunately, the extent of the bias depends itself on the true yet unknown value of $\tilde{\lambda}_{1}$. This is a challenging statistical problem, which we leave for further research. However, the result we obtained can be used to derive an upper bound on the bias. Note indeed that the bias is maximum at $\tilde{\lambda}_{1}=0$, taking a value of $1/(T-1)$. This suggests a simple rule of thumb for practitioners to quantify the expected bias they can face in the worst case scenario.

\subsection{Illustration of the theorem}

To illustrate this result, we simulate population covariance matrices $\Sigma_N(\tilde{\lambda})$ according to the moderate spike model, varying $\tilde{\lambda}_{1}$ between 0 and 1. That is, $\Sigma_N(\tilde{\lambda})=Q_N\Lambda_N(\tilde{\lambda}) Q_N^{'}$, where $\Lambda_N(\tilde{\lambda})$ is a diagonal matrix with the eigenvalues   $[aN=\tilde{\lambda}/(1-\tilde{\lambda})N, 1, \ldots,1]$ and $Q_N$ is a random orthogonal matrix. For each of the $\Sigma(\tilde{\lambda})$ covariances, we then simulate data with sample size of $T\in[4,20,100]$ and dimension
$N\in[50,200,500,1000]$ assuming a multivariate normal distribution that is i.i.d. over time. The correlation metrics being insensitive to the values of the means vector $\mu_N$, we simply set it to 0. In other words, we now simulate data from:

\begin{equation*}Y_{T,N}\stackrel{iid}{\sim} \mathcal{N}
\left(0_N, \Sigma_N(\tilde{\lambda})=Q\Lambda_N(\tilde{\lambda}) Q^{'}\right)
\end{equation*}

\begin{figure}
\caption{Theoretical bias and bias from the spiked-model simulations\label{fig:theo-simul-bias-T4}}

\begin{centering}
\includegraphics[width=0.95\columnwidth]{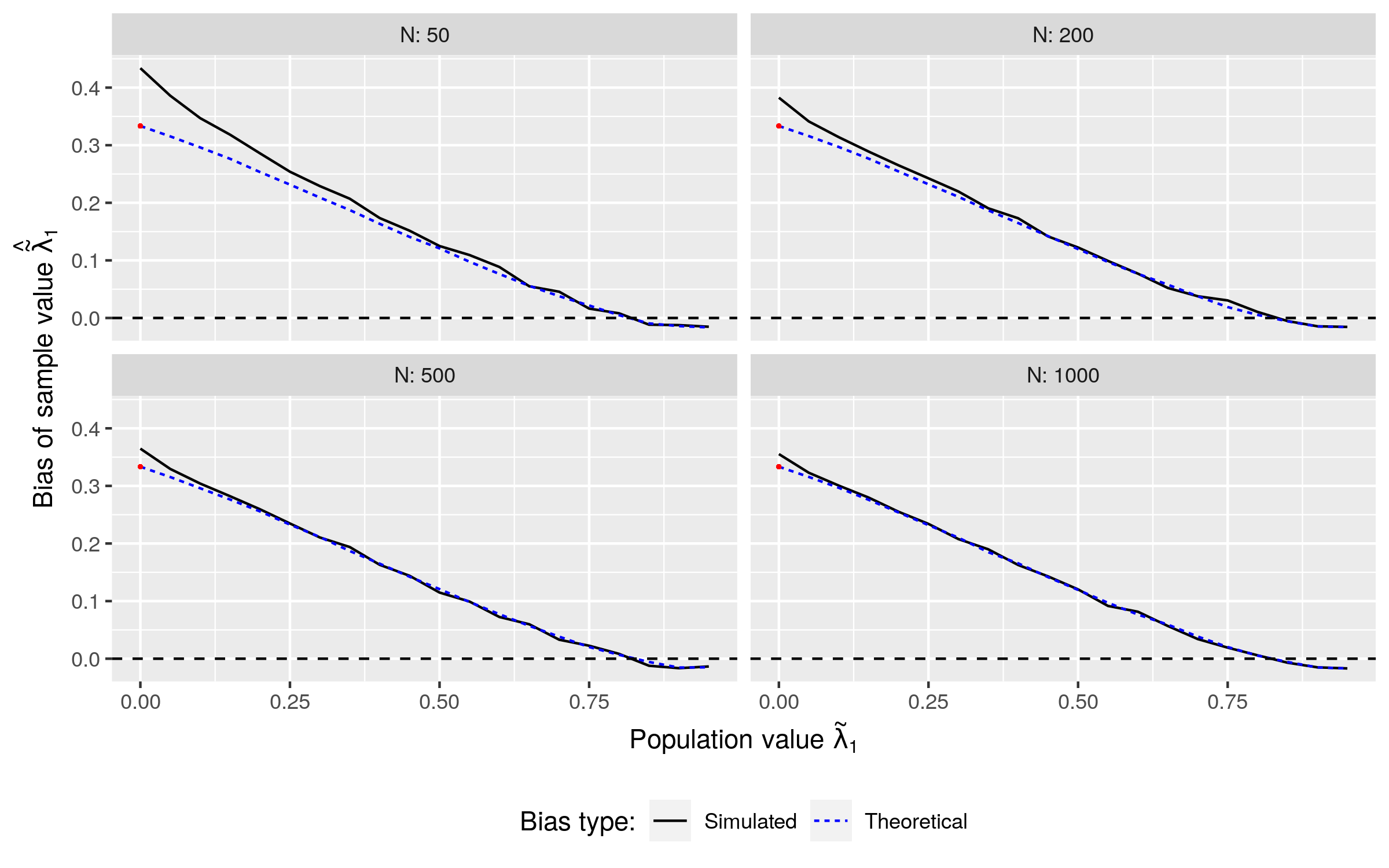}
\par\end{centering}
\end{figure}

We then estimate $\hat{\tilde{\lambda}}_{1}$ on the simulated data. Figure~\ref{fig:theo-simul-bias-T4} shows the resulting bias estimates on the y-axis, and the true population value $\tilde{\lambda}_{1}$ on the x-axis. The black line represent the bias of the values estimated on the simulated data from the spiked model, and the blue line represents the formula from Theorem~(\ref{thm:main-theorem-sample-eigenval}).
The red dot corresponds to the worst bound $1/(T-1)$. Focusing first on the behaviour of the bias, we clearly see the phenomenon observed with the SCYM-calibrated simulations shown in section~\ref{sec:Data-and-empirical-simuls}. For low values of $\tilde{\lambda}$, we observe a very substantive upwards bias, leading to over-confidence in the
quality of an insurance product. This bias decreases with increasing $\tilde{\lambda}$, and even reverses at very high values of $\tilde{\lambda}$. Comparing now the difference between the simulated bias (black line) and our asymptotic formula from (\ref{thm:main-theorem-sample-eigenval}), we see that the formula approximates remarkably well the empirical bias for dimensions as low as $N=200$. More interestingly, having a higher dimensionality $N$ is no longer a curse, but improves instead the validity of the bias formula! 

\begin{figure}
  \caption{Theoretical bias and bias from the SCYM-calibrated simulations\label{fig:SCYMul-vs-theo}}

  \begin{centering}
   \includegraphics[width=0.95\columnwidth]{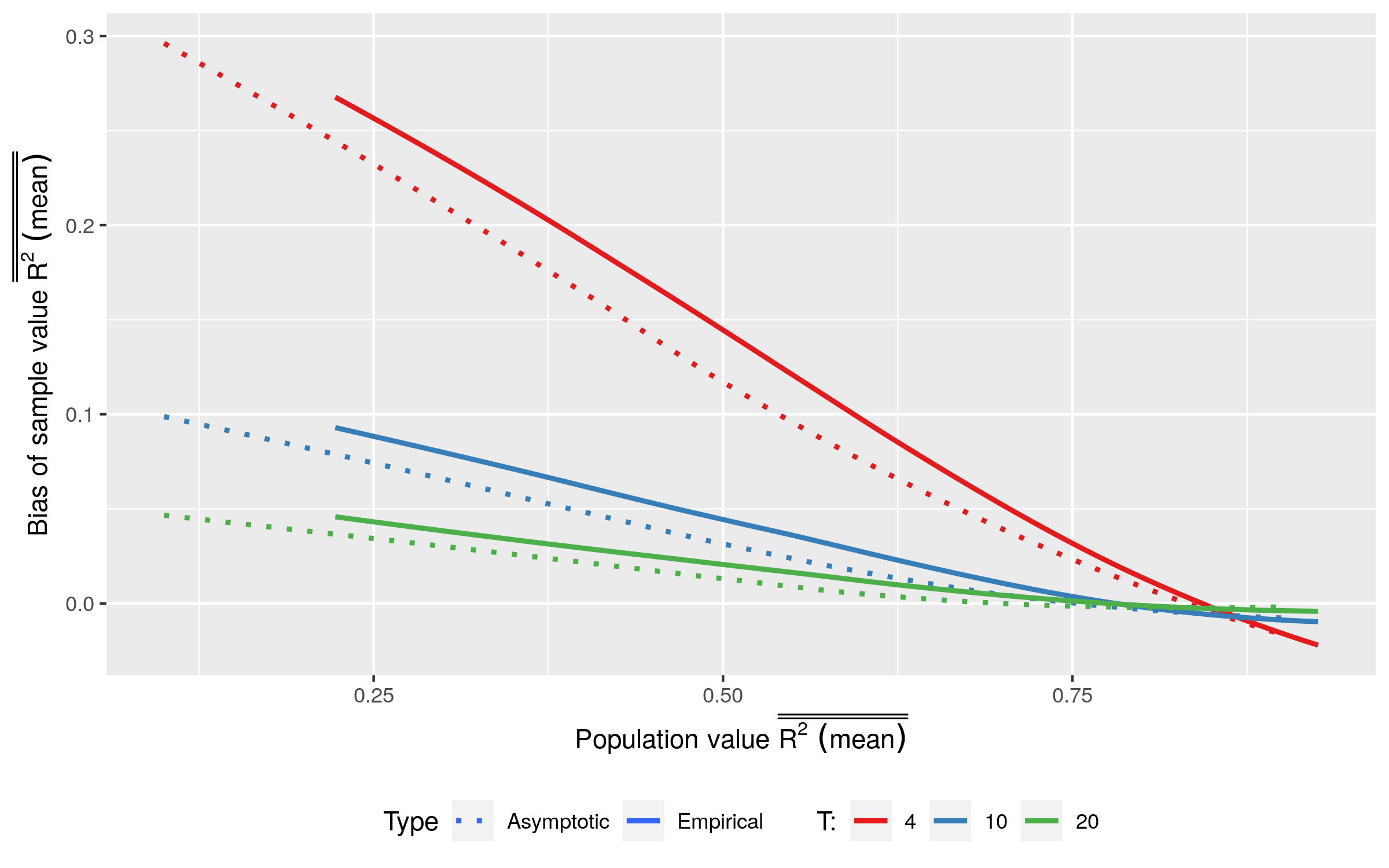}
   \end{centering}
\end{figure}

As a final test, we compare the empirical bias obtained in the calibrated simulations to the analytical bias formula. Remember that the calibrated simulations were generated based on the empirical covariance matrices estimated from the satellite data. As such, there is no guarantee that the asymptotic bias formula holds, since the latter is obtained assuming a constant spike model, which might not hold in the SCYM dataset. To further test the generality of the bias formula, we focus on the total $R^2$ measure based on the area-yield index, instead of focusing on $\hat{\tilde{\lambda}}_1=R^2(w^\star)$. In a constant spike model, one can show that these measures are very similar, but this is not necessarily the case in practice. Figure~\ref{fig:SCYMul-vs-theo} shows the empirical bias (straight line) together with the bias predicted by theorem 2 (dotted line). The analytical bias formula approximates the empirical bias well, with a difference that tends to vanish as the dimension $N$ increases. This result is very encouraging, suggesting that our choice of a constant spike model to represent the data might be relevant in practice. Furthermore, it indicates that the theory developed here provides a reliable tool for practitioners to assess the potential bias in their measures of basis risk. 

\section{Conclusion}

High resolution satellite images coupled with recent advances in machine learning are expected to yield significant progress in index insurance. However, the discussion to date has overlooked the fact that while data is becoming much richer in terms of number of fields observed, the number of years we observe remains low. As we have shown in this paper both theoretically and empirically through simulation, this introduces a downward bias in common measures of basis risk, which if left uncorrected is likely to yield overly optimistic assessments of insurance product quality.

Our paper is the first to identify this bias in commonly used measures of index insurance quality, which is important because it is likely to lead to overly optimistic assessments of product quality. The theory we develop to explain the observed bias provides a useful set of building blocks to approximate and bound the bias in real-world situations. Academics as well as organizations and governments designing new index insurance products ought to take this into account, especially when products are being designed based on a small number of time periods; the simulation methods we develop provide a strategy for generating more accurate product quality estimates.

We highlight two aspects of this bias that are especially problematic for developing country settings where satellite-based index insurance has the greatest potential benefit. First, the bias is generally larger when individual farm yields are less correlated, which is more true in rural smallholder systems than in large developed country farms. Second, the bias can actually increase as the number of fields in the data increases, meaning high resolution data may yield worse estimates of basis risk if the bias is uncorrected. For these reasons, it is critical that companies, governments, and non-governmental organizations designing index insurance products are aware of and take steps to correct the bias we study in this paper.

Future work ought to focus on developing methods to correct the bias identified in this paper. More careful modeling of the covariance between fields is a promising avenue: methods that take into account spatial correlation and/or allow the covariance matrix between fields to vary over time are promising. For example, in the worst years, when poor weather conditions are the dominant factor affecting yields, it may be that field-level outcomes are more correlated than in more typical years, when individual non-weather shocks drive most of the variation.

This paper showed evidence of bias in both linear and quantile-based measures of basis risk, yet confirmed analytically this bias only in the linear case. Focusing on linear measures of basis risk is convenient in that they are easy to understand and study analytically. At the same time, linear measures fail to capture the fact that the effect of basis risk is nonlinear: failures to accurately predict negative shocks are much more detrimental than failures in good years. Analyzing the theoretical bias of nonlinear measures of basis risk and of expected utility metrics remains a challenging task where future research will be needed. 

\appendix 
\section{Appendix}

\begin{proof}[Notation]
Let $X_n$ be a sequence of random variables with distribution function $F_n$ respectively and let $X$ be another random variable with distribution function $F$. Then, we say that $X_n$ converges in distribution to $X$, and write  $X_n\stackrel{d}{\to}X$ if for all continuity points $x$ of $F$, $F_n(x)\to F(x)$ as $n\to\infty$. We say that $X_n$ converges in probability to $X$, and write $X_n\stackrel{p}{\to}X$ if $\mathbb P(|X_n-X|>\epsilon)\to 0$ as $n\to\infty$, for any fixed $\epsilon>0$. Finally, we say that $X_n$ converges almost surely to $X$, and write $X_n\stackrel{a.s.}{\to}X$ if $\mathbb P(X_n\to X)=1$. It is well known that $X_n\stackrel{a.s.}{\to}X\implies X_n\stackrel{p}{\to} X\implies X_n\stackrel{d}{\to}X$. For details, see \citet{Billingsley1995}.
\end{proof}

\begin{proof}[Proof of Theorem \ref{thm:main-theorem-sample-eigenval}]

Let $S$ be the sample covariance matrix for the data $ Y_1,\cdots, Y_T$, i.e. $S=\tfrac{1}{T}\sum_{t=1}^T( Y_t-{\overline{Y}})( Y_t-{\overline{Y}})'$ where $\overline{Y}=\tfrac{1}{T}\sum_{t=1}^T Y_i$. Recall that $\hat{\tilde\lambda}=\lambda_{max}(S)/Trace(S)$. It is well known from standard multivariate theory (see \citet{MardiaKentEtAl1979}) that $S\stackrel{d}{=} \tfrac{1}{T}X'X=\tfrac{1}{T}\sum_{t=1}^{T-1}X_iX_i'$ where $X$ is a $(T-1)\times N$ matrix with rows $X_i\stackrel{iid}{\sim}\mathcal N_N( 0,\Sigma)$. Hence, in whatever follows, we work with the iid data $X_1,\cdots,X_{T-1}$. Also, define the dual sample covariance matrix for the data $X$, as $S_D=\tfrac{1}{T-1}XX'$.

 For the vanishing spike case, we use Theorem 1 from \citet{AhnMarronEtAl2007}, see also \citet{IshiiYataEtAl2014}.  Note that the function $A\mapsto \lambda_{max}(A)/Trace(A)$ is continuous on the space of positive definite matrices. So, we conclude that $\tfrac{\lambda_{max}(S_D)}{Trace(S_D)}\stackrel{a.s.}{\to}\tfrac{1}{T-1}$ and therefore $\tfrac{\lambda_{max}(X'X/T)}{Trace(X'X/T)}\stackrel{a.s.}{\to}\tfrac{1}{T-1}$. Due to the distributional equivalence between $S$ and $\tfrac{1}{T}X'X$ we conclude that ${\lambda_{max}(S)}/{Trace(S)}\stackrel{a.s.}{\to}\tfrac{1}{T-1}$.

For the constant and expanding spike cases, we follow the method presented in Section 4 of \citet{AhnMarronEtAl2007}. The dual covariance matrix $S_D$ can be expressed as $(T-1)S_D=aN^\alpha Z_1Z_1+b\sum_{i=2}^N Z_iZ_i'$ where $Z_i$ are iid $\mathcal N_{T-1}(0,I)$ where $I$ denotes the $(T-1)-$dimensional identity matrix. Hence, $(T-1)S_D/N^\alpha$ converges a.s. to $aZ_1Z_1'+bI$ if $\alpha=1$ (constant spike) and to $aZ_1Z_1'$ if $\alpha>1$ (expanding spike). Hence, $\lambda_{max}(S_D)/Trace(S_D)\stackrel{d}{\to}\tfrac{aC^2+b}{aC^2+b(T-1)}$ in the constant spike case and in probability to $1$ in the expanding spike case. Here, $C^2=\|Z_1\|^2\sim\chi^2_{T-1}$.
Again, utilizing the distributional equivalence between $S$ and $X'X/T$ we conclude that $\lambda_{max}(S)/Trace(S)\stackrel{d}{\to}\tfrac{aC^2+b}{aC^2+b(T-1)}$ in the moderate spike case, and $\lambda_{max}(S)/Trace(S)\stackrel{p}{\to}1$ in the expanding spike case. This completes the proof.

\end{proof}

\begin{proof}[Proof of Theorem \ref{thm:theorem-bias}]
Note that the random variables $\hat{\tilde\lambda}$ are bounded between 0 and 1. By Theorem \ref{thm:main-theorem-sample-eigenval}, in each of vanishing, constant and expanding spike cases, this sequence of random variables converges in distribution, and hence by uniform integrability, their expectations also converge to the corresponding quantities. This shows that for the vanishing and expanding spike cases, $\mathbb E(\lambda_{max}(S)/Trace(S))$ converges to $\tfrac{1}{T-1}$ and $1$ respectively. Noting that the true asymptotic largest eigenvalue shares were $0$ and $1$ respectively in the vanishing and expanding spike cases, the biases are $1/(T-1)$ and $0$ respectively.

For the constant spike case, writing $r=a/(a+b)$, we get that $$Bias(\hat{\tilde\lambda})=\mathbb E(\hat{\tilde\lambda})-r\to \mathbb E\left(\dfrac{rC^2+(1-r)}{rC^2+(1-r)(T-1)}\right)-r=\mathbb E\left(\dfrac{(1-r)(rC^2+1-r(T-1))}{rC^2+(1-r)(T-1)}\right)$$
\end{proof}

\bibliographystyle{ecta}
\bibliography{biblio_mat_synced_from_laptop, more_bib}

\end{document}